\documentclass{article}
\usepackage{spconf,amsmath,graphicx}
\usepackage{enumitem}
\usepackage{xcolor}
\usepackage{hyperref}
\usepackage{amsmath}
\setlist{nosep, leftmargin=14pt}
\usepackage{mwe}

\begin{document}

\title{Mask-guided data augmentation for multiparametric MRI generation with a rare hepatocellular carcinoma}

\name{\begin{tabular}{c} Karen Sanchez$^{\star}$$^{\diamond}$, Carlos Hinojosa$^{\star}$, Kevin Arias$^{\star}$, Henry Arguello$^{\star}$,\\ Denis Kouamé$^{\dagger}$, Olivier Meyrignac$^{\ast}$, and Adrian Basarab$^{\diamond}$ \end{tabular}}

\address{$^{\star}$HDSP research group, Universidad Industrial de Santander, Bucaramanga, 680002 Colombia.\\
$^{\dagger}$IRIT, University of Toulouse, CNRS 5505, Université Paul Sabatier Toulouse 3, France.\\
$^{\ast}$Bicêtre University Hospital, Assistance publique des hôpitaux de Paris, Biomaps, France.\\
$^{\diamond}$Univ Lyon, Université Claude Bernard Lyon 1, CNRS, Inserm, CREATIS UMR 5220, U1294, France.}

\maketitle
\begin{abstract}
Data augmentation is classically used to improve the overall performance of deep learning models. It is, however, challenging in the case of medical applications, and in particular for multiparametric datasets. For example, traditional geometric transformations used in several applications to generate synthetic images can modify in a non-realistic manner the patients' anatomy. Therefore, dedicated image generation techniques are necessary in the medical field to, for example, mimic a given pathology realistically. This paper introduces a new data augmentation architecture that generates synthetic multiparametric (T1 arterial, T1 portal, and T2) magnetic resonance images (MRI) of massive macrotrabecular subtype hepatocellular carcinoma with their corresponding tumor masks through a generative deep learning approach. The proposed architecture creates liver tumor masks and abdominal edges used as input in a Pix2Pix network for synthetic data creation. The method's efficiency is demonstrated by training it on a limited multiparametric dataset of MRI triplets from $89$ patients with liver lesions to generate $1,000$ synthetic triplets and their corresponding liver tumor masks. The resulting Frechet Inception Distance score was $86.55$. The proposed approach was among the winners of the 2021 data augmentation challenge organized by the French Society of Radiology.
\end{abstract}


\section{Introduction}
\label{sec:intro}
Massive macrotrabecular hepatocellular carcinoma (MMHCC) is a recently identified cancer subtype with high research interest due to its aggressive phenotype \cite{jeon2019macrotrabecular}. Computer-aided diagnosis, more recently combined with deep learning (DL), has emerged to support medical staff in decision-making in recent decades \cite{calderon2021bilsk, escobar2021accurate}, see, e.g., automatic tumor segmentation in magnetic resonance imaging (MRI) \cite{ranjbarzadeh2021brain}. However, DL models dedicated to healthcare still face several challenges \cite{shen2020introduction}. One of them is that their accuracy strongly depends on the availability of a large amount of labeled data \cite{sanchez2022cx}. However, the latter may be scarce, especially in the case of rare pathologies, such as MMHCC addressed in this work, isolated and generally private \cite{frid2018synthetic}.
Data augmentation has emerged as an alternative to mitigate the lack of training data. It generally generates one or more units from each input data through basic manipulations such as geometric transformations, random erasing, kernel filters, mixing images, or color space modifications \cite{chlap2021review}. Nevertheless, while traditional manipulations are efficient in natural image-related applications, they do not perform similarly well in medical applications \cite{maldjian2007approach}. For instance, X-rays of the abdomen or thorax flipped horizontally generate images with situs inversus condition, a genetic malformation characterized by the inverted position of the organs \cite{maldjian2007approach}.
Furthermore, current algorithms for data augmentation do not explicitly guarantee the global structure of the image nor the interrelation of its elements, main requirements that in turn may limit the variability of new examples \cite{wymer2020phase}. Finally, data augmentation strategies are even more scarce for multiparametric medical imaging, where multiple images of a patient are expected to reflect the same medical condition \cite{mule2020multiphase}. In this context, the contributions of this work are the following:
    \noindent (\textbf{i}) a new data augmentation strategy that generates multiparametric MRIs, and can preserve the global structure and interrelationship between tissues in the data distribution;\\
    \noindent (\textbf{ii}) a strategy to generate the corresponding liver tumor segmentation mask of each synthetic patient;\\
    \noindent (\textbf{iii}) an original approach to synthesize the massive macrotrabecular tumor in multiparametric MRI.
To assess the effectiveness and potential of the proposed framework, extensive simulations were performed on a private dataset of $89$ real patients created by the French Society of Radiology for a 2021 challenge on this topic. $1,000$ MMHCC synthetic multiparameter MRI cases and their segmentation masks were generated. The diversity and fidelity of the synthetic images were evaluated qualitatively by three radiologists using the Likert scale \cite{mule2022generative}, and quantitatively using the Frechet Inception Distance metric. The results were compared with existing approaches such as Pix2Pix \cite{isola2017image}, CycleGAN \cite{zhu2017unpaired} and the proposed strategy by changing the generative backbone.

\section{Synthetic MRI generation method}
This section describes the proposed data augmentation strategy for MMHCC MRI. The method generates three different images, T1-weighted with fat saturation (T1-FS) arterial phase, T1-FS portal phase, and T2-weighted, and also a common segmentation MMHCC mask. The overall method is depicted in Fig. \ref{fig:framework}. It contains two main stages. First, it creates new tumor and surrounding tissue edge masks as described in Section \ref{masks}. Second, it trains a Pix2Pix generative adversarial network to produce synthetic multiparametric triplets of MMHCC MRI from the masks generated at the previous stage, as detailed in Section \ref{pix2pix}.
\vspace{-0.1cm}
\subsection{First stage: Generation of input masks}
\label{masks}
\vspace{-0.1cm}
Highlighted as the first stage in Fig. \ref{fig:framework}, the objective of this approach is to create valid anatomical borders and tumor locations. As illustrated in the left-hand side of Fig. \ref{fig:framework}, its main steps are as follows.

\noindent (\textbf{i}) \textbf{Geometric transformations of tumor masks}. Let us denote by $X1_m, X2_m, X3_m$ the training data available, \textit{i.e.}, the three MRIs available for each patient affected by MMHCC. Furthermore, let us denote the tumor segmentation masks of the training dataset as ${X}_m$, for $m = 1...M$, where $M$ is the total number of patients in the training database. Let $s = 1...S$ be the subscript of synthetic cases generated from each patient, where $S$ is their total number. First, each ${X}_m$ is modified with multiple-random geometric operations, such as zoom, rotation, flip, and translation to generate new tumor segmentation masks denoted as $\bar{X}_{ms}$ and named ``Output tumor mask'' in Fig \ref{fig:framework}. In this step, the translation operation for creating new masks is spatially limited to ensure that the centroid of the synthetic tumor is inscribed in the region that can anatomically correspond to the liver of an average person, avoiding situs inversus. Note that this step provides the first guarantee for data augmentation, since in the synthetic images to be generated the position of the tumor will be determined by the mask created at this stage.\\
\noindent (\textbf{ii}) \textbf{Manual liver segmentation}. Second, an experienced radiologist manually segmented the liver on all $X1_m, X2_m, X3_m$ MRIs using a freely available computer vision annotation tool (CVAT). Output liver segmentation images are denoted as $L_m$.\\
\noindent (\textbf{iii}) \textbf{Algorithm for edge detection.} Third, the proposed method uses the traditional Canny edge detector algorithm \cite{rong2014improved} to estimate all the contours inside $X1_m, X2_m,$ and $X3_m$ images. In this step, the method normalizes the input image pixels between 0 and 255. Then, the Canny operator uses a multi-stage algorithm to detect a wide range of edges in the images with a randomly set of $P$ thresholds between $30$ and $120$, a heuristically defined range. Each contour image, denoted by $E_{mp}$, will be used to delimit the distribution of organs and tissues in the synthetic images. Since the threshold to calculate the borders is variable, this method can generate multiple contour images from the same MRI, which is the second guarantee of data augmentation.\\
\noindent (\textbf{iv})  \textbf{Tumor and liver intersection mask.} This fourth step aims to ensure that the synthetic MRIs generated have a tumor located strictly inside the liver. For this purpose, the proposed method calculates an intersection image, denoted as $I_s$, between the tumor $\bar{X}_{ms}$ and the liver masks $L_m$, i.e., $I_{ms} = \bar{X}_{ms} \cap {L}_m$. \\
\noindent (\textbf{v}) \textbf{New tumor and edges mask.} The last step of the first stage overlays the tumor masks $I_{ms}$ from the previous step on the edge detection images $E_{mp}$. Let us denote these new images as $N_{msp}$, which will be used as input for the adversarial generative network of the second stage.\\
\vspace{-0.75cm}
\subsection{Second stage: Synthetic MRI generation}
\label{pix2pix}
\begin{figure*}[t]
	\centering
	\includegraphics[width=\linewidth]{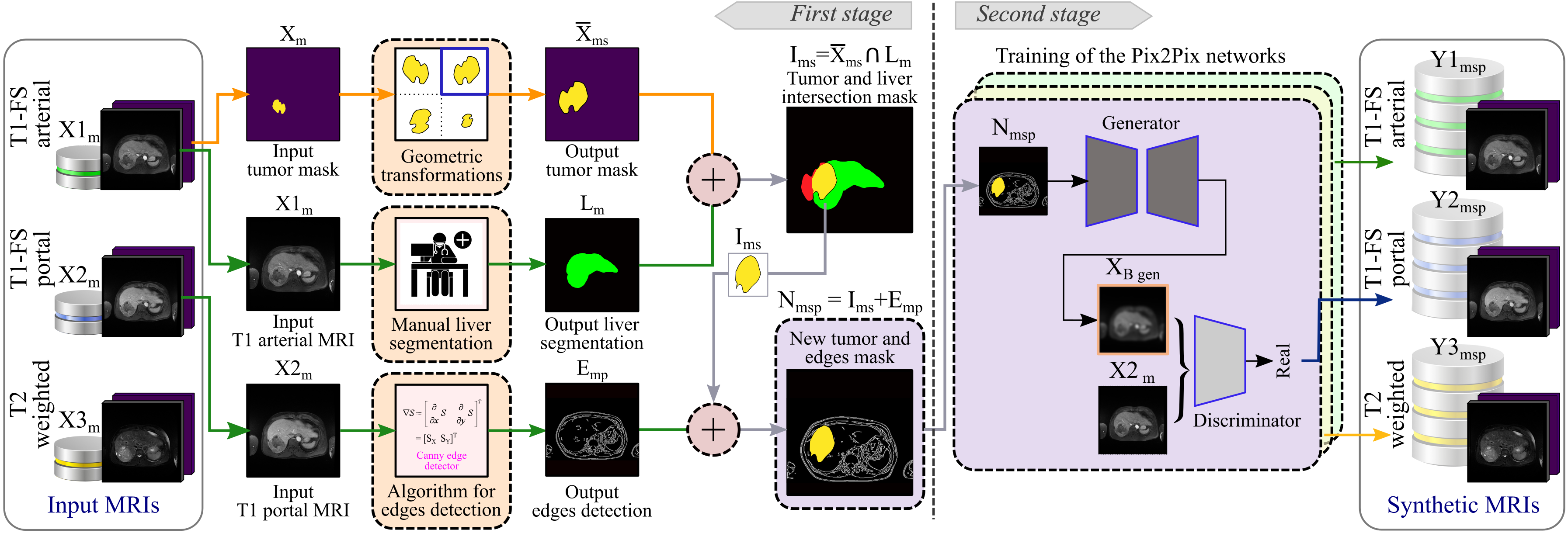}
	\caption{General architecture of the proposed data augmentation method for multiparametric MRIs of MMHCC. The method is composed of two main stages divided in six steps. From left to right, it first creates multiple $\bar{X}_{ms}$ geometric transformations of each ${X}_m$ tumor mask from the training dataset. Second, a radiologist performed manual liver segmentation $L_m$ on training MRIs. Next, the Canny edge detector algorithm calculates the edges $E_{mp}$ from ${X_m}$. New tumor masks $I_{msp}$ are created by the intersection of the output tumor $\bar{X}_{ms}$ images and the liver segmentation masks $L_m$. Then, tumor and edges images $N_{msp}$ are created by assembling the tumor masks $I_{ms}$ and the contour images $E_{mp}$. Finally, three Pix2pix networks are customized and trained to create three multiparametric images ${Y1}_{msp}, {Y2}_{msp}, {Y3}_{msp}$, from the conditions imposed by the image $N_{msp}$.}
	\label{fig:framework}
\end{figure*}

\vspace{-0.1cm}
\noindent (\textbf{vi}) \textbf{Generative adversarial network:} As illustrated on the right side of Fig. \ref{fig:framework}, the Second stage covers the generation of synthetic images through an adversarial network. The data augmentation strategy proposed in this paper uses the Pix2Pix \cite{isola2017image} architecture, a network where the output image generation is conditional on an input one, \textit{i.e.}, image-to-image translation.
Specifically, three Pix2Pix networks are configured in parallel, the three generators are trained to receive as input the same image $N_{msp}$ resulting from the first stage, and each produces a different type of MRI acquisition (T1-FS arterial, T1-FS portal, and T2) as output. On the other hand, the discriminator part of each Pix2Pix network is fed with both the output images of the generator (target) and the real training set (source) to determine through the loss function if the target is a plausible transformation of the source image. Therefore, a different set of training data ($X1_m, X2_m,$ or $X3_m$) is used as source for each Pix2Pix. The GAN that receives the $X1_m$ images can generate the synthetic $Y1_{msp}$. Similarly, networks trained with the images $X2_m$ and $X3_m$ will produce $Y2_{msp}$ and $Y3_{msp}$, respectively. Note that the $N_{msp}$ images were designed to delimit the distribution of organs, tissues, and the location of MMHCC tumors on the synthetic images. This guarantees an adequate anatomical composition of the new MRI, which traditional GANs do not condition.

\section{Experiments and Results}

\subsection{Training dataset, implementation details and evaluation metric}
\label{sec:data}
\textit{\textbf{Dataset.}} The proposed approach was evaluated through a private dataset of $267$ multiparametric MRIs corresponding to $89$ patients with MMHCC, made available by the French Society of Radiology within the ``GAN-based data augmentation of rare liver cancers: The SFR 2021 Artificial Intelligence Data Challenge'' \cite{mule2022generative}. The objective of this challenge was to design deep learning methods capable of generating any number of synthetic MRI triplets ($1,000$ have been targeted by the challenge) from these available MR images. Specifically, for each image, a tumor segmentation mask and three acquisitions were available: T1-FS arterial, T1-FS portal, and T2-weighted. \\
\textit{\textbf{Implementation details.}} The implementation of the proposed method was performed with Python using PyTorch. The Pix2Pix networks were trained using the Adam solver with 200 epochs, a learning rate of 0.0001, a batch size of 24, and momentum parameters $\beta_1 = 0.5, \beta_2 = 0.9$. All MRI were resized to $256 \times 256$ pixels for the training process.\\
\textit{\textbf{Evaluation metric.}} The quality of the generated synthetic images was evaluated through the Frechet Inception Distance (FID) score \cite{heusel2017gans}. FID is a well-known metric to evaluate the quality of generated images, quantifying the distance between original and synthetic images created by a model generator. Mathematically, the quality assessment of the generated images can be expressed as follows: 
\vspace{-0.2cm}
\begin{equation} 
    \ell_{FID} = \Vert \mu_{R} - \mu_{F}\Vert^{2} - Tr\left( \Sigma_{R} + \Sigma_{F} - 2 \Sigma_{R} \Sigma_{F} \right),
    \label{FID_metric}
\end{equation}
\noindent where $Tr \left( \cdot \right)$ is the trace of the argument matrix, the pair $\left(R,F \right)$ denotes the distributions of the real and generated data, $\left( \mu_{R}, \mu_{F} \right)$ represents the mean of each distribution, and $\left( \Sigma_{R}, \Sigma_{F} \right)$ indicates their covariance matrices, respectively. A lower FID score indicates better quality images, while a high FID stands for lower quality synthetic images in a nearly linear relationship with those in the training dataset.

\subsection{Qualitative results}

\begin{figure}[h]
	\centering
	\vspace{-0.15cm}
	\includegraphics[width=0.99\linewidth]{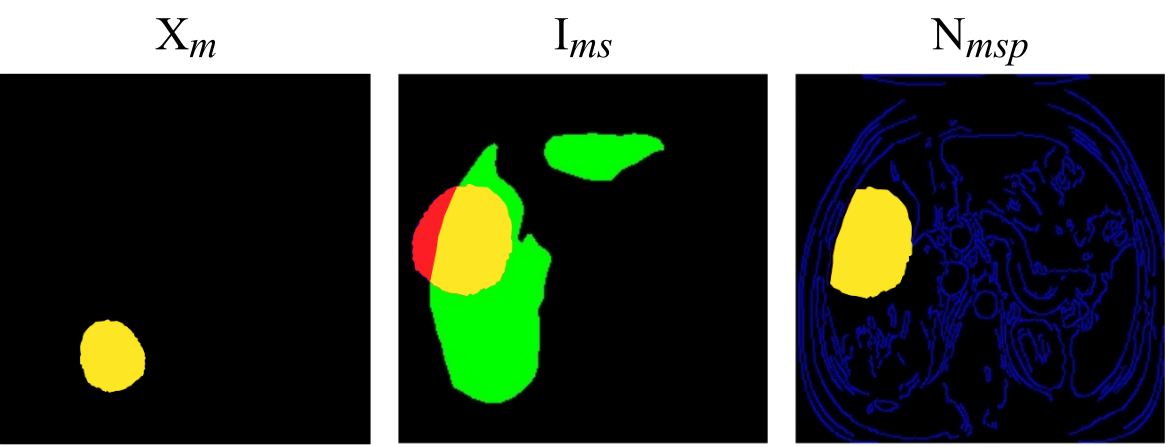}
	\caption{Visual results of a randomly chosen sample of the first stage of the proposed framework. From left to right: an original tumor mask from the training set, the intersection between the modified tumor and the liver mask, and the anatomical liver contour with the tumor mask.}
	\label{fig:masks}
\end{figure}

This subsection presents the visual results of the proposed framework. Fig. \ref{fig:masks} illustrates a randomly chosen sample result of the first stage: a real tumor mask ${X}_m$, its transformation and intersection with the liver mask ${I}_{ms}$, and the overlay of the tumor and the contour image ${N}_{msp}$. Note that, in this example, the tumor mask ${X}_m$ was horizontally inverted, enlarged, and moved to a different location to generate the transformed tumor mask $\bar{X}_{ms}$. 
The center image of Fig. \ref{fig:masks} shows the result of an intersection mask ${I}_{ms}$, where the liver manual segmentation $L_m$ is in green, the section of the tumor located inside the liver in yellow, and the tumor outside the liver in red. In the right-hand side of Fig. \ref{fig:masks}, the intersecting tumor region $I_{ms}$ is overlaid with an edge image $E_{mp}$ to generate a $N_{msp}$ image.

\begin{figure}
    \centering
    \includegraphics[width=0.84\linewidth]{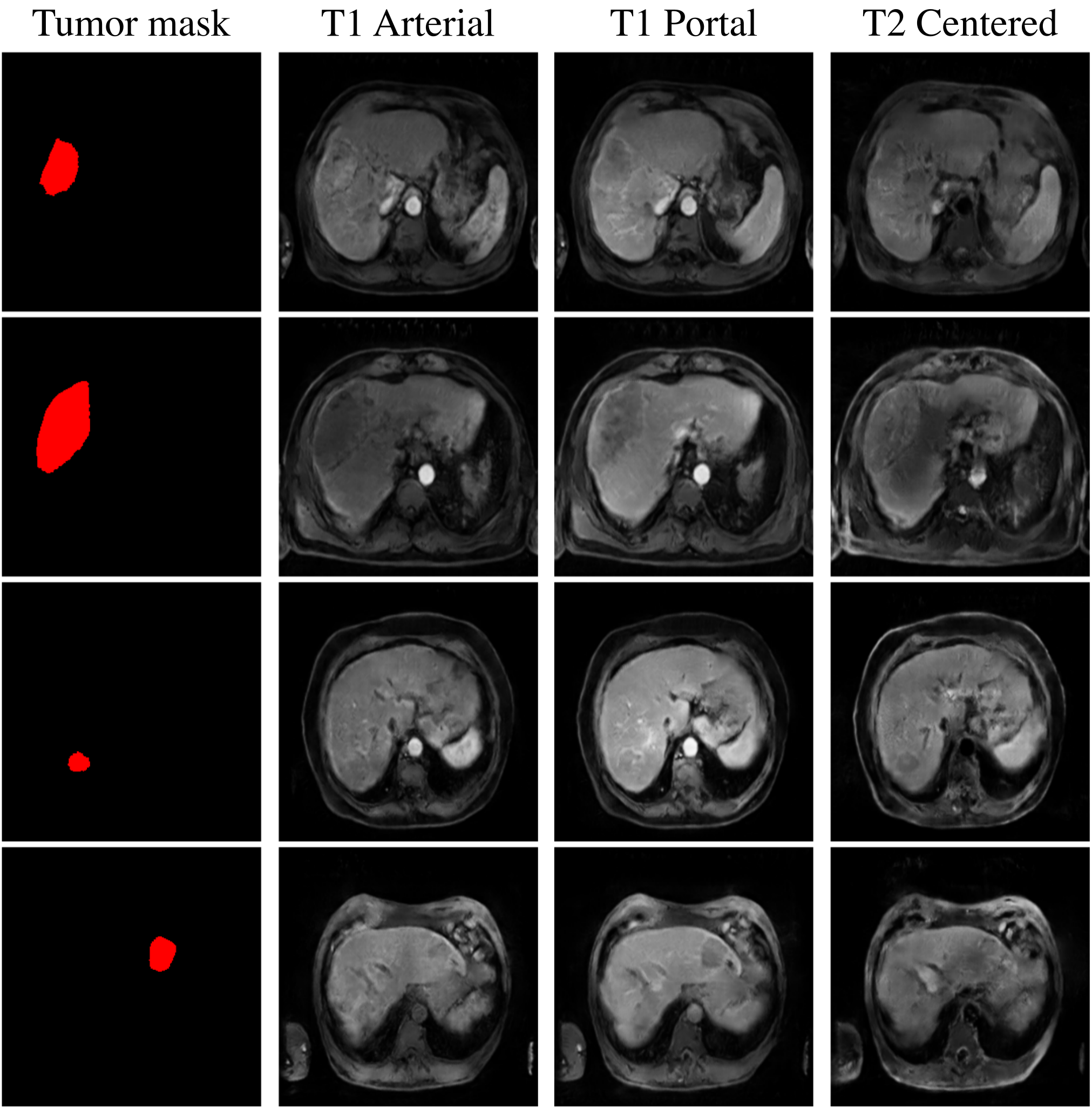}
    \caption{Visual results of the second stage. Generated synthetic images for each type of parametric and phase image, for three virtual patients.}
    \label{fig:mri_res}
\end{figure}

\newpage
The proposed method was tested to generate $1,000$ synthetic patient cases, each with three multiparametric MRIs, i.e., $3,000$ images and their associated MMHCC masks. Figure \ref{fig:mri_res} illustrates four random cases of synthetic patients, one per row, generated by the proposed method. The first column refers to the tumor segmentation mask ($I_{ms}$), the second column to the T1-FS arterial $Y1_{msp}$, the third corresponds to the T1-FS portal $Y2_{msp}$, and the last column to the T2-weighted $Y3_{msp}$. More results are available on the project website \footnote{\url{https://github.com/carlosh93/SFR\_HB-IRIT-UIS}}, allowing to appreciate the diversity and fidelity of the synthetic MRIs.

\subsection{Quantitative results}
\vspace{-0.2cm}
In this subsection, the effectiveness of the proposed data augmentation framework is evaluated quantitatively compared to state-of-the-art methods. All the methods were trained on the dataset described in Section \ref{sec:data} and used to generate $1,000$ synthetic multiparametric cases of MMHCC MRIs.
Table \ref{table1} regroups quantitative results with four methods: a Pix2Pix network \cite{isola2017image}, a CycleGAN \cite{zhu2017unpaired}, the proposed method, using a CycleGAN in the second stage and the proposed method using Pix2Pix as described in this paper.
FID scores for each type of acquisition are provided in Table \ref{table1} for each MRI type individually and for all the synthetic MRIs in the last column. As explained in \ref{FID_metric}, FID is calculated from the distance between the training and the generated data. Therefore, $89$ real and $1,000$ synthetic MRIs of each MRI type are used to calculate the results for the first three columns. The score in the last column results from the total number of $267$ real images and $3,000$ synthetic ones. The best result is shown in bold, and the second-best is underlined. As mentioned in the Subsection \ref{sec:data}, low FID score results are desired, corresponding to higher quality synthetic images. Note that the strategy proposed herein achieves the best results for each individual MRI class and overall. The same framework with CycleGAN instead of Pix2Pix provides the second-best results. Well-known state-of-the-art GANs such as Pix2Pix and CycleGAN provide higher FID, thus less realistic synthetic MRIs. Finally, another major drawback of traditional GANs compared to the proposed framework is that they do not provide the tumor segmentation mask corresponding to the synthetic images.

\begin{table}
\caption{Quantitative MRI generation results (FID score) for each MRI type and overall.}
\centering
\small
\begin{tabular}{l|ccc|c}
\hline
\multicolumn{5}{c}{($\downarrow$) Frechet Inception Distance score (FID)}\\ \hline
\begin{tabular}[c]{@{}l@{}}\textbf{MRI type $\rightarrow$}\\\textbf{$\downarrow$ Method} \end{tabular} & {\textbf{T1 Arterial}} & {\textbf{T1 Portal}} & {\textbf{T2}} & {\textbf{All}}\\ \hline
Pix2Pix \cite{isola2017image}   & 274.87  & 284.95 & 267.77 & 234.39  \\
CycleGAN {\cite{zhu2017unpaired}} & 251.12 & 241.27 &  258.33 & 210.32  \\ 
\begin{tabular}[c]{@{}l@{}}Proposed method\\ with CycleGAN\end{tabular} & \underline{226.76}    & \underline{211.32} & \underline{229.22}  & \underline{193.12}\\
Proposed method & \textbf{115.13} & \textbf{102.47} & \textbf{123.37} & \textbf{86.55} \\ \hline
\end{tabular}
\label{table1}
\end{table}

\section{Conclusion}

This paper presented a novel framework for synthetic multiparametric MRI generation, \textit{i.e.}, with two different sequence types (T1-FS and T2) and three different acquisitions (arterial, portal, and weighted) associated with patients with a specific rare liver tumor. The key idea was to generate new tumor masks through the intersection between liver segmentation and augmented tumor masks. This step was crucial to guarantee the anatomically correct position of the tumors in the generated images. The new tumor masks overlapped with edges of real abdominal MRI were used to initialize three Pix2Pix networks trained to produce three related multiparametric MRIs corresponding to a new synthetic patient. Our method can generate an unlimited number of cases, since it can generate infinite segmentation masks by combining different thresholds for Canny-based edge detection and multiple geometric transformations of the tumor masks. The simulation results showed an appropriate FID score and promising visual results. In the future, we plan to investigate and validate how the quality of the generated images through the proposed method improves the final behaviour of specific computer vision tasks, such as segmentation or classification. 

\section{Compliance with Ethical Standards}
This is a numerical simulation study for which no ethical approval was required.

\section{Acknowledgments}
The authors thank the French Society of Radiology for the labeled and anonymized magnetic resonance images used in this study. In addition, the authors thank Minciencias through the CTO 910-2019 project ECOSNORD, and the Vicerrectoría de Investigación y Extensión UIS project code 3735. The authors have no relevant financial or non-financial interests to disclose. 

\bibliographystyle{IEEEbib}
\bibliography{strings,refs}

\begin{thebibliography}{10}

\bibitem{jeon2019macrotrabecular}
Yejoo Jeon, Mark Benedict, Tamar Taddei, Dhanpat Jain, and Xuchen Zhang,
\newblock ``Macrotrabecular hepatocellular carcinoma,''
\newblock {\em The American Journal of Surgical Pathology}, vol. 43, no. 7, pp.
  943--948, 2019.

\bibitem{calderon2021bilsk}
Camilo Calder{\'o}n, Karen Sanchez, Sergio Castillo, and Henry Arguello,
\newblock ``Bilsk: A bilinear convolutional neural network approach for skin
  lesion classification,''
\newblock {\em Computer Methods and Programs in Biomedicine Update}, vol. 1,
  pp. 100036, 2021.

\bibitem{escobar2021accurate}
Jessica Escobar, Karen Sanchez, Carlos Hinojosa, Henry Arguello, and Sergio
  Castillo,
\newblock ``Accurate deep learning-based gastrointestinal disease
  classification via transfer learning strategy,''
\newblock in {\em 2021 XXIII Symposium on Image, Signal Processing and
  Artificial Vision (STSIVA)}. IEEE, 2021, pp. 1--5.

\bibitem{ranjbarzadeh2021brain}
Ramin Ranjbarzadeh, Abbas Bagherian~Kasgari, Saeid Jafarzadeh~Ghoushchi,
  Shokofeh Anari, Maryam Naseri, and Malika Bendechache,
\newblock ``Brain tumor segmentation based on deep learning and an attention
  mechanism using mri multi-modalities brain images,''
\newblock {\em Scientific Reports}, vol. 11, no. 1, pp. 1--17, 2021.

\bibitem{shen2020introduction}
Chenyang Shen, Dan Nguyen, Zhiguo Zhou, Steve~B Jiang, Bin Dong, and Xun Jia,
\newblock ``An introduction to deep learning in medical physics: advantages,
  potential, and challenges,''
\newblock {\em Physics in Medicine \& Biology}, vol. 65, no. 5, pp. 05TR01,
  2020.

\bibitem{sanchez2022cx}
Karen Sanchez, Carlos Hinojosa, Henry Arguello, Denis Kouam{\'e}, Olivier
  Meyrignac, and Adrian Basarab,
\newblock ``Cx-dagan: Domain adaptation for pneumonia diagnosis on a small
  chest x-ray dataset,''
\newblock {\em IEEE Transactions on Medical Imaging}, 2022.

\bibitem{frid2018synthetic}
Maayan Frid-Adar, Eyal Klang, Michal Amitai, Jacob Goldberger, and Hayit
  Greenspan,
\newblock ``Synthetic data augmentation using gan for improved liver lesion
  classification,''
\newblock in {\em 2018 IEEE 15th international symposium on biomedical imaging
  (ISBI 2018)}. IEEE, 2018, pp. 289--293.

\bibitem{chlap2021review}
Phillip Chlap, Hang Min, Nym Vandenberg, Jason Dowling, Lois Holloway, and
  Annette Haworth,
\newblock ``A review of medical image data augmentation techniques for deep
  learning applications,''
\newblock {\em Journal of Medical Imaging and Radiation Oncology}, vol. 65, no.
  5, pp. 545--563, 2021.

\bibitem{maldjian2007approach}
Pierre~D Maldjian and Muhamed Saric,
\newblock ``Approach to dextrocardia in adults,''
\newblock {\em American Journal of Roentgenology}, vol. 188, no. 6\_supplement,
  pp. S39--S49, 2007.

\bibitem{wymer2020phase}
David~T Wymer, Kunal~P Patel, William~F Burke~III, and Vinay~K Bhatia,
\newblock ``Phase-contrast mri: physics, techniques, and clinical
  applications,''
\newblock {\em Radiographics}, vol. 40, no. 1, pp. 122--140, 2020.

\bibitem{mule2020multiphase}
S{\'e}bastien Mul{\'e}, Athena Galletto~Pregliasco, Arthur Tenenhaus, Rym
  Kharrat, Giuliana Amaddeo, Laurence Baranes, Alexis Laurent, H{\'e}l{\`e}ne
  Regnault, Daniele Sommacale, Marjane Djabbari, et~al.,
\newblock ``Multiphase liver mri for identifying the macrotrabecular-massive
  subtype of hepatocellular carcinoma,''
\newblock {\em Radiology}, vol. 295, no. 3, pp. 562--571, 2020.

\bibitem{mule2022generative}
S{\'e}bastien Mul{\'e}, Littisha Lawrance, Younes Belkouchi, Val{\'e}rie
  Vilgrain, Mait{\'e} Lewin, Herv{\'e} Trillaud, Christine Hoeffel, Val{\'e}rie
  Laurent, Samy Ammari, Eric Morand, et~al.,
\newblock ``Generative adversarial networks (gan)-based data augmentation of
  rare liver cancers: The sfr 2021 artificial intelligence data challenge,''
\newblock {\em Diagnostic and Interventional Imaging}, 2022.

\bibitem{isola2017image}
Phillip Isola, Jun-Yan Zhu, Tinghui Zhou, and Alexei~A Efros,
\newblock ``Image-to-image translation with conditional adversarial networks,''
\newblock in {\em Proceedings of the IEEE conference on computer vision and
  pattern recognition}, 2017, pp. 1125--1134.

\bibitem{zhu2017unpaired}
Jun-Yan Zhu, Taesung Park, Phillip Isola, and Alexei~A Efros,
\newblock ``Unpaired image-to-image translation using cycle-consistent
  adversarial networks,''
\newblock in {\em Proceedings of the IEEE international conference on computer
  vision}, 2017, pp. 2223--2232.

\bibitem{rong2014improved}
Weibin Rong, Zhanjing Li, Wei Zhang, and Lining Sun,
\newblock ``An improved canny edge detection algorithm,''
\newblock in {\em 2014 IEEE international conference on mechatronics and
  automation}. IEEE, 2014, pp. 577--582.

\bibitem{heusel2017gans}
Martin Heusel, Hubert Ramsauer, Thomas Unterthiner, Bernhard Nessler, and Sepp
  Hochreiter,
\newblock ``Gans trained by a two time-scale update rule converge to a local
  nash equilibrium,''
\newblock {\em Advances in neural information processing systems}, vol. 30,
  2017.

\end{thebibliography}

\end{document}